\documentclass[a4paper,11pt]{article}
\usepackage{pos_nologo}
\usepackage{physics}
\usepackage{comment}
\usepackage{wrapfig}
\usepackage{here}

\title{Lattice Weyl Fermion on a Single Spherical Domain-Wall}

\author*[a]{Shoto~Aoki}
\author[b]{Hidenori~Fukaya}
\author*[b]{Naoto~Kan}

\affiliation[a]{Graduate School of Arts and Sciences, University of Tokyo Komaba, \\
Meguro-ku, Tokyo 1538902, Japan}

\affiliation[b]{Department of Physics, Osaka University,\\
  Toyonaka, Osaka 560-0043, Japan}

\emailAdd{shotoaoki@g.ecc.u-tokyo.ac.jp}
\emailAdd{hfukaya@het.phys.sci.osaka-u.ac.jp}
\emailAdd{naotokan000@gmail.com}

\abstract{
We investigate a single spherical domain-wall embedded in a three-dimensional Euclidean lattice.  
We employ the Shamir-type domain-wall fermion formulation, where the negative mass region is confined inside the $S^2$ domain-wall, while the positive mass outside is taken to the infinite limit, allowing the exterior region to be neglected effectively. 
In the absence of a gauge potential, Weyl fermions emerge as edge-localized modes along the $S^2$ domain-wall. 
When a nontrivial $U(1)$ gauge potential is present, additional zero modes with opposite chirality appears, localized at the center near the monopole. 
This centrally localized mode originates from the effective mass contribution of the Wilson term, which induces a domain-wall near the monopole. 
\par
Preprint number: UT-Komaba/25-2, OU-HET-1264
}

\FullConference{The 41st International Symposium on Lattice Field Theory (LATTICE2024)\\
 28 July - 3 August 2024\\
Liverpool, UK\\}

\begin{document}
\maketitle

\section{Introduction}
In the conventional flat domain-wall fermion framework \cite{Kaplan:1992bt,Shamir:1993zy,Furman:1994ky}, two domain walls are required on a five-dimensional lattice to localize massless left- and right-handed fermions. This setup has been explored as a potential foundation for constructing chiral gauge theories on the lattice. 
However, decoupling unwanted modes from gauge fields presents significant challenges \cite{Okumura:2016dsr}. For instance, the cobordism invariance of instanton backgrounds ensures that both domain-walls host an equal number of zero modes, making it difficult to isolate purely chiral dynamics in the low-energy theory. 
Recently, symmetric mass generation mechanism is studied to gap out these unwanted modes without breaking chiral symmetry \cite{Wen:2013ppa}. 
However, this method relies on highly nonperturbative dynamics, which remain challenging to fully understand and implement.

When the domain-wall is allowed to have nontrivial curvature, it does not have to appear in left and right pairs.
In particular, when the domain-wall is a closed manifold like a sphere, it can be embedded into a one-dimension higher flat space without any other domain-walls.
Nash embedding theorems \cite{Nash1956} guarantee that every curved manifold can be isometrically embedded into a finite-dimensional Euclidean spaces.
Moreover, the metric and spin connection are uniquely determined up to the local Lorentz transformation by the embedding.

Two of the authors have embodied this concept in lattice gauge theory \cite{Aoki:2022cwg,Aoki:2022aez}.
They have confirmed that the spherical domain-wall fermion systems on a periodic flat lattice hold edge-localized modes, which feel gravity through the induced spin connection.
The obtained eigenvalue spectrum is consistent with the continuum theory prediction.
Moreover, automatic recovery of the rotational invariance is also observed.

Recently, a similar system was discussed in Refs.~\cite{Kaplan:2023pxd,Kaplan:2023pvd,Clancy:2024bjb,Kaplan:2024ezz}.
Since the single spherical domain-wall fermion can have an isolated left- or right-handed edge fermion, they argued that chiral gauge theory can be formulated on the lattice.
Although this proposal is appealing in the free fermion case, it is nontrivial to reproduce the Weyl fermion anomaly and its cancellation.
In particular, it was pointed out in Ref.~\cite{Golterman:2024ccm} that an axial current, which should be anomalous, can become conserved, which contradicts with the continuum theory.

In this work, we consider a two-dimensional spherical domain-wall embedded in a three-dimensional square lattice. The result was already published in Ref.~\cite{Aoki:2024bwx}.
As is expected, the free fermion Dirac operator spectrum is consistent with that of a single Weyl fermion.
The gravitational effect through the induced spin connection is confirmed in the gap of the eigenvalues.
However, when a topologically nontrivial $U(1)$ gauge background is given, we find an additional zero mode localized at the center of the sphere, where the gauge field becomes singular.
We identify this additional zero mode as an edge-localized state on a small but finite domain-wall near the center, which is dynamically created by the Wilson term\footnote{In Ref.~\cite{Aoki:2023lqp}, we discussed that the same mechanism can explain the Witten effect, in which the monopole background captures an electron zero mode and becomes a dyon.}.

\section{Without gauge fields}
\label{sec: without gauge}
In this paper, we consider Dirac operators
\begin{align}
  D= & \sum_i\sigma^i \qty(\pdv{}{x^i}-iA_i)+m(r)
  \label{eq:DiracOp}
\end{align} 
on a three-dimensional Euclidean space $\mathbb{R}^3$. Here, $\sigma_i$ is a Pauli matrix, $A_i$ is a $U(1)$ gauge field in the $i$-th direction, which will be considered in a later section, and $m(r)$ is a mass term. To put an $S^2$ domain-wall with the radius $r_0$, the form of $m(r)$ is
\begin{align}
  m(r)= \begin{cases}
    -m & (r \leq r_0) \\
    +M & (r>r_0)
  \end{cases} ,
\end{align}
where $m$ and $M$ are positive constants. We denote the negative mass region as $\mathbb{D}^3$. Taking the limit of $M\to +\infty$. It is equivalent to ignoring the outside of $\mathbb{D}^3$ and imposing a boundary condition 
\begin{align}
  \sigma_r \psi(r_0) =+ \psi(r_0),
\end{align}
where the chirality operator $\sigma_r$ is given by
\begin{align}
  \sigma_r= \sum_i \sigma^i x^i/r.
\end{align}
Then, we obtain a Shamir domain-wall fermion system. If we employ the Hermitian conjugate operator $D^\dagger$, we obtain a boundary condition with $\sigma_r=-1$.

We first address the free fermion case with $\textbf{A}=0$. Since the system is rotational symmetric, the Dirac operator $D$ commutes with the total angular momentum operators
\begin{align}
  J_a= L_a + \frac{1}{2} \sigma_a, 
\end{align}
where $L_a$ is the orbital contribution
\begin{align}
  L_a= -i \epsilon_{abc} x^b \pdv{}{x^c},
\end{align}
and $ \frac{1}{2} \sigma_a$ is the spin part. Let $j$ be the total angular momentum quantum number. Then, we have
\begin{align}
  j= \frac{1}{2},~\frac{3}{2},~ \frac{5}{2},\cdots.
\end{align}

Respecting the rotational symmetry, we rewrite $D$ as
\begin{align}
  D= \sigma_r \qty(\pdv{}{r} +\frac{1}{r} -\frac{iD^{S^2}}{r} ) -m,
\end{align}
where $D^{S^2}$ is an effective Dirac operator given by
\begin{align}
  iD^{S^2}=\sigma^a \qty(L_a )+1=\sigma^a J_a -\frac{1}{2}.
\end{align}
$D^{S^2}$ and $\sigma_r$ commute with the total angular momentum operator $J_a $, but anti-commute with each other. Since the square of $D^{S^2}$ is give by
\begin{align}
  (iD^{S^2})^2 = \mathbf{J}^2 +\frac{1}{4},  
\end{align}
the eigenvalues of $D^{S^2}$ are not zero. We find the spinors $\chi_{j,j_3, \pm}$ which satisfy
\begin{align}
  \mathbf{J}^2 \chi_{j,j_3, \pm}&= j(j+1)\chi_{j,j_3, \pm},\\
  J_3\chi_{j,j_3, \pm}&= j_3 \chi_{j,j_3,\pm}\quad (j_3=-j,-j+1,\cdots , j-1,j),\\
  iD^{S^2} \chi_{j,j_3, \pm} &= \pm \nu \chi_{j,j_3,\pm }\quad (\nu=  j+\frac{1}{2}),\\
  \sigma_r \chi_{j,j_3,\pm}&= \chi_{j,j_3, \mp}.
\end{align}

The eigenstate of $D^\dagger D \psi =E^2 \psi $ and $D D^\dagger  \psi =E^2 \psi$ are expanded by $\chi_{j,j_3,\pm}$. Assuming that $E$ is positive, we obtain the edge localized modes as
\begin{align}\label{eq:DW localized mode}
  \psi_{j,j_3}^{D^\dagger D}&= \frac{1}{\sqrt{r}} \qty(  \sqrt{m-E} I_{\nu-1/2}(\sqrt{m^2-E^2}r) \chi_{j,j_3,+} +  \sqrt{m+E} I_{\nu+1/2}(\sqrt{m^2-E^2}r)\chi_{j,j_3,-} ) \\
  \psi_{j,j_3}^{DD^\dagger }&= \frac{1}{\sqrt{r}} \qty(  \sqrt{m-E} I_{\nu-1/2}(\sqrt{m^2-E^2}r) \chi_{j,j_3,+} -  \sqrt{m+E} I_{\nu+1/2}(\sqrt{m^2-E^2}r)\chi_{j,j_3,-} )
\end{align} 
for $E<m$. They have a $(2j+1)$-fold degeneracy. From the boundary condition for $D$ and $D^\dagger$, $E$ must satisfies
\begin{align}\label{eq:E continuum}
    \frac{ I_{\nu-1/2}(\sqrt{m^2-E^2}r_0) }{ I_{\nu+1/2}(\sqrt{m^2-E^2}r_0)}= \frac{ \sqrt{m+E} }{ \sqrt{m-E}}.
\end{align}
In the large mass limit, $E$ converges to
\begin{align}
  E \to \frac{\nu}{ r_0}.
\end{align}
Thus, identifying $\sigma_r$ as the chirality operator, the Weyl fermions appear at the wall, on which the effective massless Dirac operator $iD^{S^2}$ acts. 

Let us rewrite $D^{S^2}$ in a standard form of the Dirac operator on $S^2$. Taking the gauge transformation or local Lorentz transformation, 
\begin{align}
  R(\theta, \phi)= \exp(i \frac{\theta}{2} \sigma_2 ) \exp(i \frac{\phi}{2} \sigma_3 ),
\end{align}
we obtain
\begin{align}
  e^{-i \phi/2}R(\theta, \phi) iD^{S^2} R(\theta,\phi)^{-1} e^{i \phi/2}= - \sigma_3 \qty(\sigma_1 \pdv{}{\theta} +\frac{\sigma_2}{\sin \theta } \qty(  \pdv{}{\phi}-i \qty(-\frac{1}{2} +\frac{\cos\theta}{2 } \sigma_3 ) ) ).
  \label{eq:S2DiracNoGauge}
\end{align}
The transformation by $e^{i\phi/2}$ makes $R(\theta,\phi)$ periodic in $\phi \to \phi +2\pi$. Then, $- \frac{1}{2}$ next to $\pdv{}{\phi}$ is generated as a spin$^c$ connection and the wave function becomes $2\pi$-periodic with respect to $\phi$. The last term $\frac{\cos\theta}{2 } \sigma_3 $ is the spin connection on $S^2$, which is related to the gravitational effect.

\begin{wrapfigure}[16]{r}[0pt]{0.4\textwidth}
  \centering
  \includegraphics[bb= 0 0 273 271, width =0.4\textwidth]{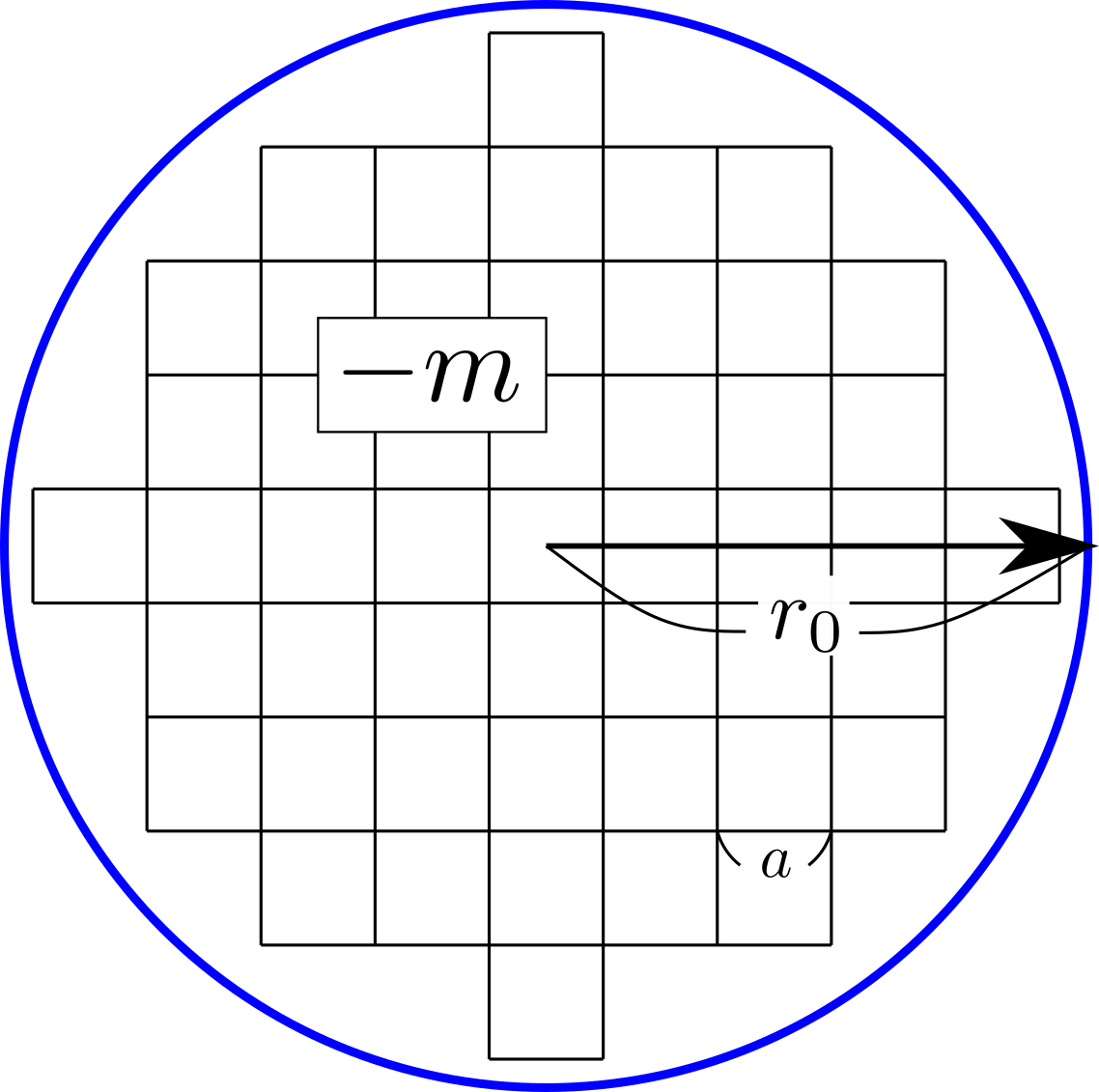}
  \caption{Two-dimensional disk with the radius $r_0$ and lattice spacing $a$. We assign the negative mass $-m<0$ on the disk.}
  \label{fig:ShamirTypeDW}
\end{wrapfigure}

Next, we show that the Weyl fermion system appear at the wall on a lattice space. We discretize the three-dimensional disk $\mathbb{D}^3$ with the radius $r_0$ and lattice spacing $a$. We depict the two-dimensional version of the disk in Fig. \ref{fig:ShamirTypeDW}. Let $x=(x^1,x^2, x^3)$ denote the lattice point, which takes
\begin{align}
  x^{i}=\cdots,~-\frac{3}{2}a,~-\frac{1}{2}a,~\frac{1}{2}a,~\frac{3}{2}a, \cdots,
\end{align}
and satisfies $r^2=\sum_{i}(x^i)^2\leq r_0^2$. The forward difference and its conjugate are given by
\begin{align}
	\label{eq:diff_op_free}
  \nabla_i \psi(x)= \psi(x+ a\hat{i}) -\psi(x), \\
  \nabla_i^\dagger \psi(x)= \psi(x- a\hat{i}) -\psi(x).
\end{align}
Here, $\hat{i}$ is a unit vector in the $i$-th direction. If the length of $x\pm a \hat{i}$ exceeds $r_0$, then $\psi(x\pm a \hat{i})=0$. As well as the continuum theory, we have the Wilson Dirac operator,
\begin{align}\label{eq:Wilson Dirac op on S2}
  D_W =\frac{1}{a}\qty(\sum_{i=1}^3\qty[\sigma_i\frac{\nabla_i-\nabla^\dagger_i}{2} +\frac{1}{2}\nabla_i \nabla^\dagger_i ]- am ).
\end{align}
We also define the chirality operator 
\begin{align}
  \sigma_r(x)= \frac{1}{r}(\sigma^1 x^1+\sigma^2 x^2+\sigma^3 x^3 ).
\end{align}

\begin{figure}[t]
  \centering
  \includegraphics[width=\textwidth,bb=0 0 1008 432]{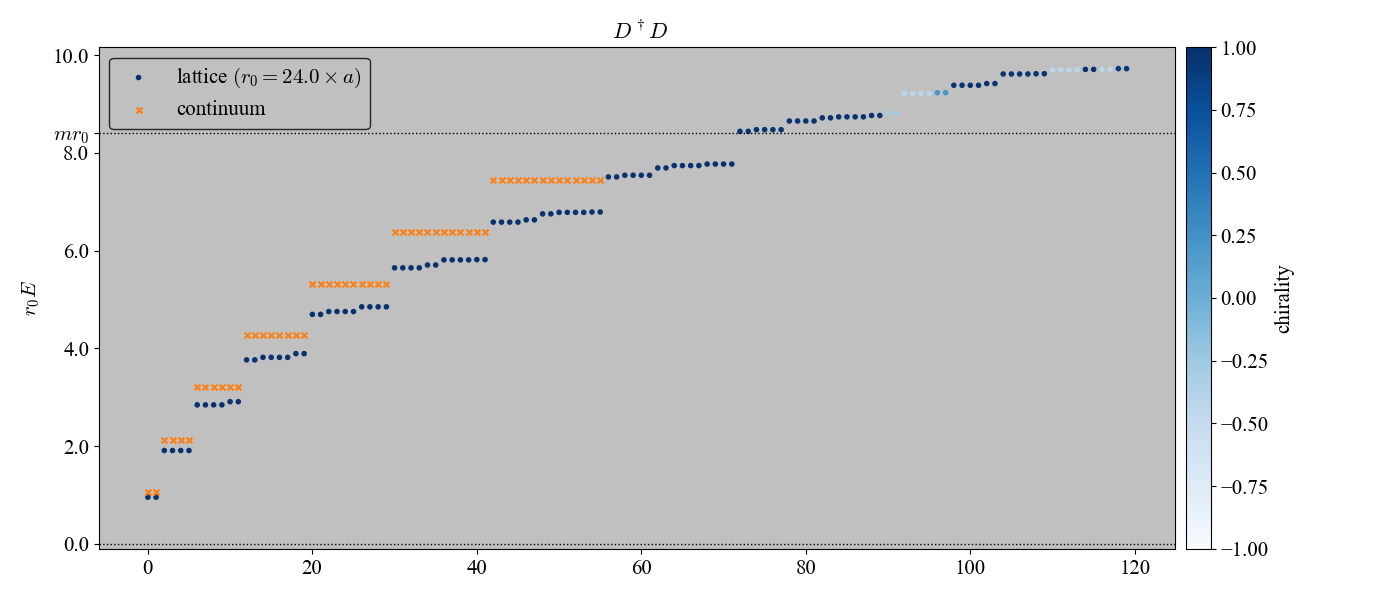}
   \caption{The spectrum of $D_W^\dagger D_W$ at $r_0=24a$ and $ma=0.35~(mr_0=8.4)$. We plot $E r_0$, which is the square root of the eigenvalue normalized by $r_0$. The color gradation represents the chirality, which is the expectation value of $\sigma_r$. $mr_0$ is shown by the dotted line.}
   \label{fig: Spectrum of DdaggerD}
\end{figure}

We numerically solve the eigenvalue problem $D_W^\dagger D_W \psi=E^2 \psi $ at $r_0=24a $ and $ma=0.35$. We plot numerical results for $E r_0$ by filled symbols in Fig \ref{fig: Spectrum of DdaggerD}, which is the square root of the eigenvalue normalized by $r_0$. The color gradation is the expectation value of $\sigma_r$. We can find that all modes below $mr_0$, which is represented by the dotted line, have the positive chirality. The lattice data agrees well with the continuum prediction (Eq.~\eqref{eq:E continuum}) designated by the orange cross symbols. The finite mass and $r_0$ slightly shift $Er_0$ from $\nu=+1,~2,\cdots$. The degeneracy looks $(2j+1)$-fold, but we can see the misalignment due to the violation of the continuous rotational symmetry. 

\begin{figure}[t]
\centering
\includegraphics[width=\textwidth,bb=0 0 1008 432]{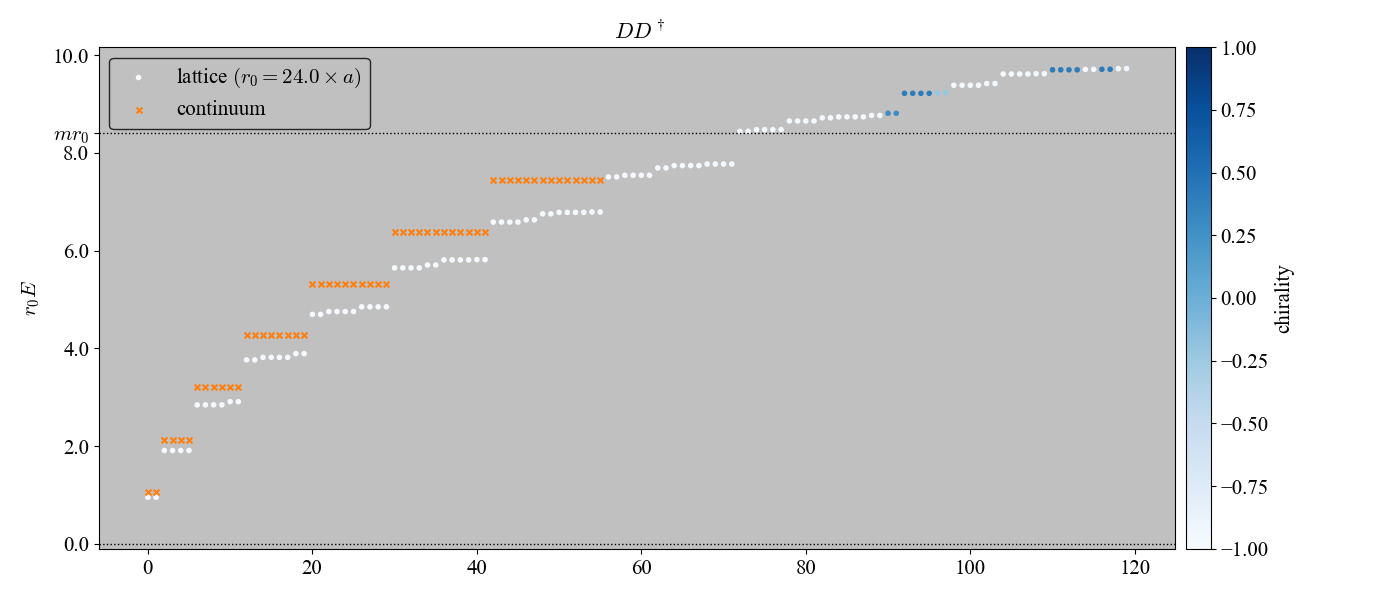}
 \caption{The spectrum of $D_W D_W^\dagger$. The parameter is the same as Fig.~\ref{fig: Spectrum of DdaggerD}.}
 \label{fig: Spectrum of DDdagger}
\end{figure}

The gap from zero is interpreted as the gravitational effect on $S^2$. In general, the eigenvalues $\lambda$ of the Dirac operator on an $n$-dimensional compact manifold satisfies 
\begin{align}
  \lambda^2 \geq \frac{n}{4(n-1)} \min_{x\in M} R(x),
\end{align}
where $R$ is the scalar curvature of the manifold \cite{Friedrich1980}. In our case, the scalar curvature is a constant function: $R=n(n-1)=2$, and the eigenvalues are larger than one. We can see this gap in Fig.~\ref{fig: Spectrum of DdaggerD} and Fig.~\ref{fig: Spectrum of DDdagger}.

We also plot the eigenvalue of $D_W D_W^\dagger$ in Fig. \ref{fig: Spectrum of DDdagger}. The spectrum is the same as Fig \ref{fig: Spectrum of DdaggerD}, but the chirality is opposite. 

\begin{figure}
\centering
\includegraphics[width=\textwidth,bb=0 0 905 291]{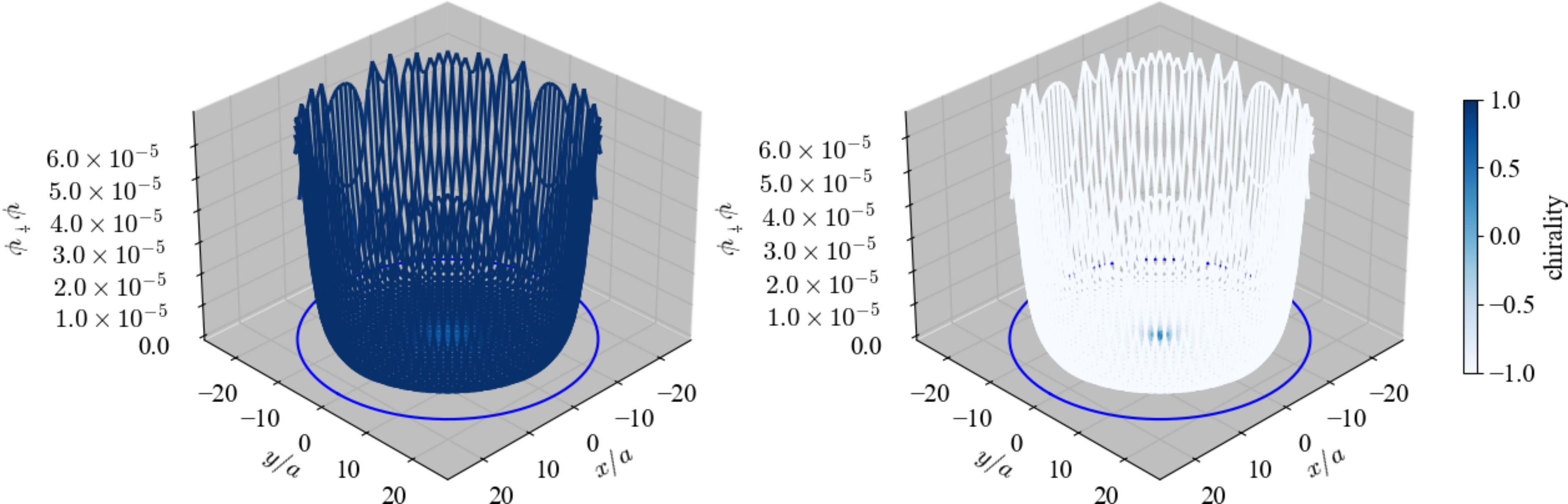}
  \caption{The amplitude of the lowest eigenstate of $D_W D_W^\dagger$ (left panel) and that of $ D_W^\dagger D_W$ (right panel) at $r_0=24a$ and $ma=0.35$ at a slice $x^3=a/2$. The color gradation is defined by Eq.~\eqref{eq: point-wise chirality}.
  } 
  \label{fig: amplitude}
\end{figure}

In Fig.~\ref{fig: amplitude}, we show the amplitude distribution of the lowest mode of $D_W^\dagger D_W$ and $D_W D_W^\dagger $ at a slice $x^3=a/2$. The color gradation represents the point-wise chirality,
\begin{align}
  \label{eq: point-wise chirality}
  \psi_k (x)^\dagger \sigma_r(x) \psi_k (x)/ \psi_k (x)^\dagger \psi_k (x),
\end{align}
where $k$ is the label of the eigenstates. We find that these modes are localized at the domain-wall.

\begin{figure}[h]
\centering
\includegraphics[width=0.8\textwidth,bb= 0 0 461 346]{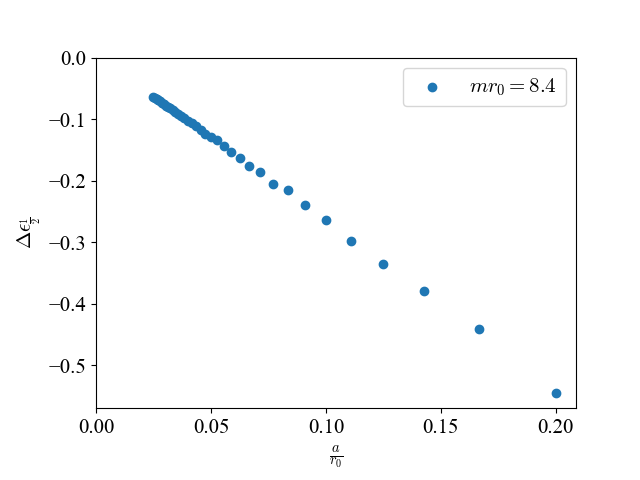}
  \caption{The relative error of $E_{1/2}$ plotted as the function of the lattice spacing $a$ normalized by $r_0$. We fix $mr_0=8.4$.}
  \label{fig: continuum limit}
\end{figure}

To evaluate the systematic error due to the lattice space $a$, we compute the eigenvalues by changing the lattice space $a$ while keeping $mr_0= 8.4$. We plot the relative deviation of the lowest eigenvalue,
\begin{align}
  \Delta \epsilon_{1/2}= \frac{ {E_{1/2} - E_{1/2}^{\text{cont}}}} {E_{1/2} ^{\text{cont}}},
\end{align}
in Fig.~\ref{fig: continuum limit}. Here, the subscript $1/2$ means the total angular momentum quantum number $j=1/2$. We can see that our lattice results linearly converge to the continuum predictions in the limit of $a\to 0$. The finite volume effect is also discussed in Ref.~\cite{Aoki:2024bwx}.

\begin{figure}
\centering
\includegraphics[width=0.8\textwidth,bb=0 0 461 346]{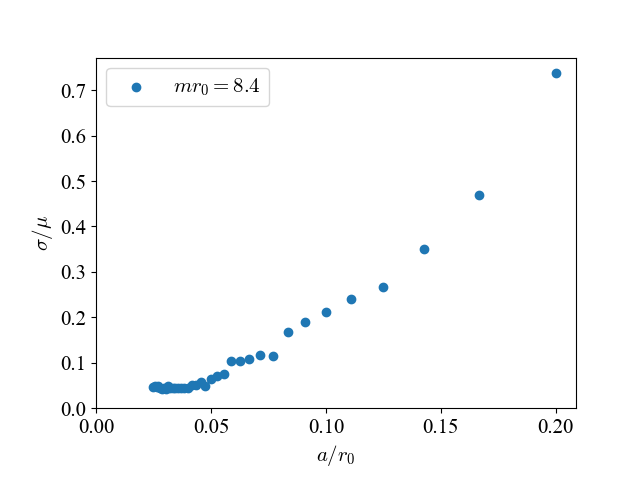}
  \caption{The standard deviation $\sigma$ normalized by the average of peaks at the boundary plotted as the function of the lattice spacing $a$. Here, we keep $mr_0=8.4$.}
  \label{fig: continuumlimit_rot}
\end{figure}

Finally, we consider the restoration of the rotational symmetry in the continuum limit. In Fig.~\ref{fig: amplitude}, there are spiky peaks, which violate the rotational symmetry. To measure the violation, we take the standard deviation $\sigma$ of the peaks near boundary normalized by their average $\mu$ in Fig. \ref{fig: continuumlimit_rot}. We can see that our results converge to zero in $a \to 0$. This indicates that the rotational symmetry automatically recovers in the continuum limit. 

Thus, we conclude that the Weyl fermion appear at the boundary in the absence of the monopole even in the lattice theory. Apparently, there exists no mirror fermion with the opposite chirality. This single domain-wall fermion system may be useful for constructing chiral gauge theories on the lattice. Although, the chirality is not perfect unless $mr_0 \to 0$.

\section{With gauge fields}
\label{sec: with gauge}
In the previous section, we discussed the Shamir domain-wall fermion system in the absence of a gauge potential.
In this section, let us turn on a nontrivial gauge potential
\begin{align}
	A_x=\frac{-q_m y}{2r(r+z)}, \quad A_y=\frac{q_m x}{2r(r+z)}, \quad A_z=0,
	\label{eq:GaugePotential}
\end{align}
where the monopole charge $q_m$ is constrained to be an integer by Dirac's quantization condition.
The gauge configuration \eqref{eq:GaugePotential} has the Dirac string at $x=y=0$ and $z<0$.
Later, we will demonstrate that the Dirac string has no physical effects. 
In the continuum discussion, we ignore it for simplicity.

In the presence of a nontrivial gauge potential, the angular momentum operator is modified to 
\begin{align}
	L_a=-i \epsilon_{abc}x^b \left({\partial \over \partial x^c}-iA_c\right)-{n x_a \over 2r},
\end{align}
where we denote the quantized monopole charge as $q_m=n\in \mathbb{Z}$.
With this modification, the total angular momentum $j$ becomes 
	\begin{align}
		j=\frac{|n|-1}{2}, \frac{|n|-1}{2}+1, \frac{|n|-1}{2}+2,\ldots
	\end{align}
under the condition $j \ge 0$.
We also note that the Dirac operator on the $S^2$ domain-wall (which reduces to Eq.~\eqref{eq:S2DiracNoGauge} when the gauge field vanishes) is given by
\begin{align}
	-\sigma_3&\left[\sigma_1\left({\partial \over \partial \theta}-i\hat{A}_\theta\right)+{\sigma_2 \over \sin\theta}\left({\partial \over \partial \phi}-i\hat{A}_\phi+{i \over 2}-{i \cos\theta \over 2}\sigma_3\right)\right] \notag \\
	&=-\sigma_3\left[\sigma_1{\partial \over \partial \theta}+{\sigma_2 \over \sin\theta}\left({\partial \over \partial \phi}-i{n\over 2}\frac{\sin^2\theta}{1+\cos\theta}+{i \over 2}-{i \cos\theta \over 2}\sigma_3\right)\right],
\end{align}
where $\hat{A}_\theta$ and $\hat{A}_\phi$ are the dimensionless gauge potential given by $\hat{A}_\theta=r A_\theta$ and $\hat{A}_\phi=r\sin\theta A_\phi$, respectively.

When the total angular momentum is $j=(|n|-1)/2$ with $|n|\ge 1$, we have a solution that did not exist in the absence of the gauge potential.
We denote an eigenstate of $\mathbf{J}^2, J_3, iD^{S^2}$ and $\sigma_r$ satisfying
\begin{align}
  \mathbf{J}^2 \chi_{j,j_3, 0}&= j(j+1)\chi_{j,j_3, 0},\\
  J_3\chi_{j,j_3, 0}&= j_3 \chi_{j,j_3,0}, \\
  iD^{S^2} \chi_{j,j_3, 0} &= 0,~\\
  \sigma_r \chi_{j,j_3,0}&={\rm sign}(n)\, \chi_{j,j_3, 0},
  \label{eq:chiral_chi_0}
\end{align}
as $\chi_{j,j_3, 0}$.
Then we find an edge-localized solution
	\begin{align}
		\psi_0=\frac{1}{\sqrt{r}}f(r)\chi_{j,j_3, 0}
	\end{align}
with
	\begin{align}
		f(r)=C I_{1/2}(\kappa r)=C\sqrt{\frac{2}{\pi \kappa r}}\sinh(\kappa r),
	\end{align}
where $C$ is a normalization constant.

For $n<0$, it follows that from Eq.~\eqref{eq:chiral_chi_0} that this solution cannot satisfy the boundary condition with a positive sign $\sigma_r \psi_0(r=r_0)=+\psi_0(r=r_0)$.
This fact affects the index of the Dirac operator $D^{S^2}$ on the sphere. From the Atiyah--Singer index theorem \cite{Atiyah:1963zz}, we obtain
\begin{align}
	{\rm Index}(iD^{S^2})=\frac{1}{2\pi}\int_{S^2} F=n.
\end{align}
This implies that the Weyl fermion with positive chirality does not contribute negatively to the index when $n<0$.
We also note that the boundary condition $\sigma_r D\psi_0(r=r_0)=-D\psi_0(r=r_0)$ is not satisfied except in the case of $D\psi_0(r=r_0)=0$ when taking the limit of $m\to \infty$.
Therefore, we have no chiral zero mode both of $D^\dagger D$ and $iD^{S^2}$ in the presence of the nontrivial gauge field with $n<0$ except for $m\to\infty$.

Next, let us move on to the lattice analysis.
In lattice gauge theory, the $U(1)$ gauge potential is introduced by the link variable 
\begin{align}
	U_i(x)=\exp\left(i\int_x^{x+\hat{i}} A_i(x')dx'^i\right),
	\label{eq:link_variable}
\end{align}
where the vector potential $A_i(x)$ is given by Eq.~\eqref{eq:GaugePotential}.
The covariant difference operators are given by
\begin{align}
	\nabla_i\psi(x)=U_i(x)\psi(x+a\hat{i})-\psi(x).
\end{align}

As mentioned earlier, it can be verified that the Dirac string at $x=y=0, z<0$ has no physical effect under the link variable Eq.~\eqref{eq:link_variable}.
Let $p_\pm$ be the values of the plaquettes in $xy$-plane, centered at $x_\pm=(0,0,\pm a/2)$.
These two values are computed as
\begin{align}
	p_\pm&=\exp\bigg(i\int_{-a/2}^{a/2}dx\, A_1(x,-a/2,\pm a/2) + \int_{-a/2}^{a/2}dy\, A_2(a/2,y,\pm a/2) \notag \\
		&\qquad \qquad -\int_{-a/2}^{a/2}dx\, A_1(x,a/2,\pm a/2) -\int_{-a/2}^{a/2}dy\, A_2(-a/2,y,\pm a/2) \bigg) \\
		&=e^{\pm in {\pi \over 3} }.
\end{align}
We see that the result is symmetric, which implies that Dirac string intersecting the plaquette at $x_-$ does not influence the value $p_-$.
Thus, we conclude that the Dirac string has no physical effects.

\begin{figure}
\centering
\includegraphics[width=\textwidth, bb=0 0 1008 829]{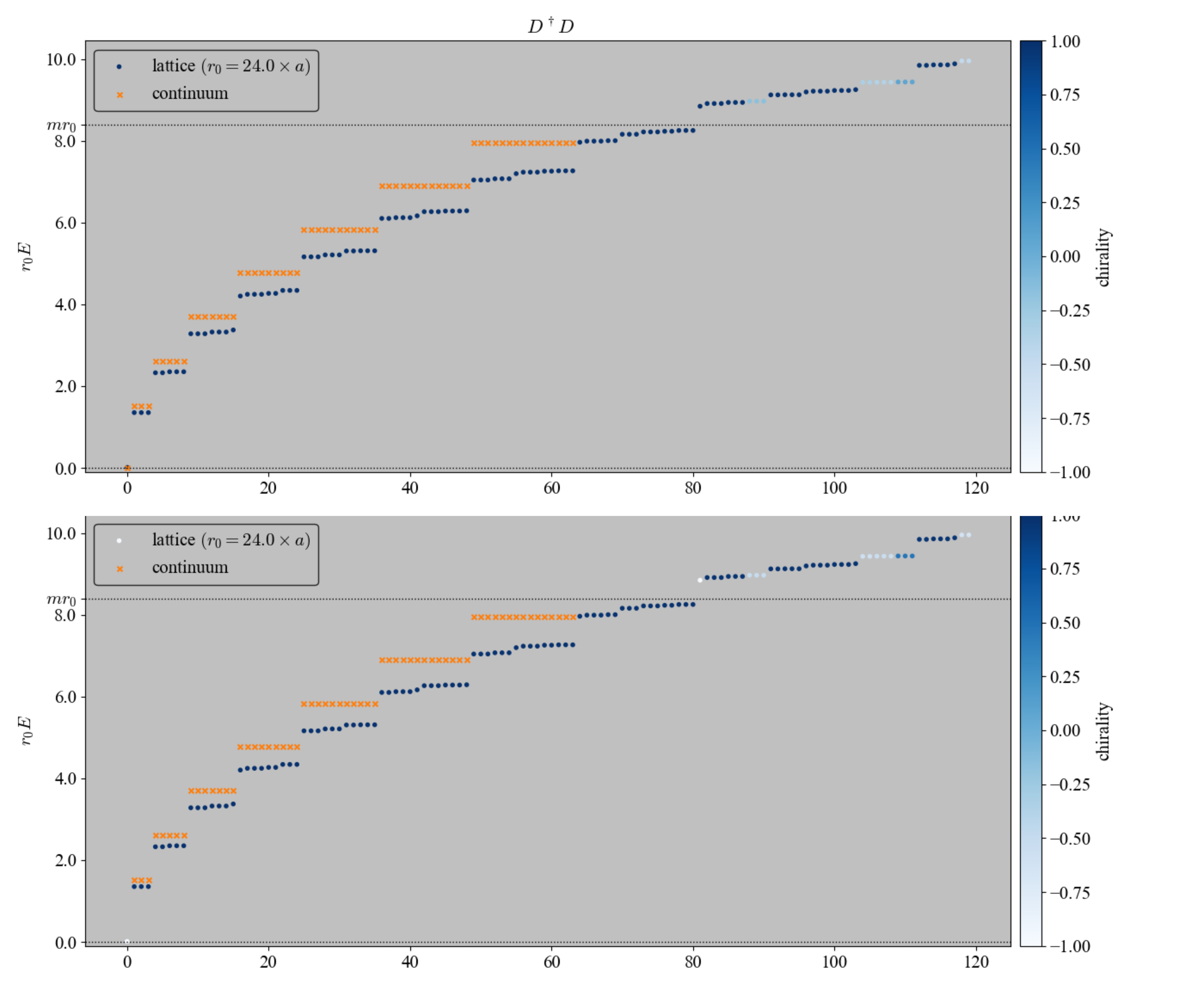}
  \caption{The plot of the eigenvalues of $D_W^\dagger D_W$ with $n=1$ (upper panel) and $n=-1$ (lower panel). }
  \label{fig: spectrum n=1}
\end{figure}

\begin{figure}
\centering
\includegraphics[width=0.9\textwidth, bb= 0 0 494 291]{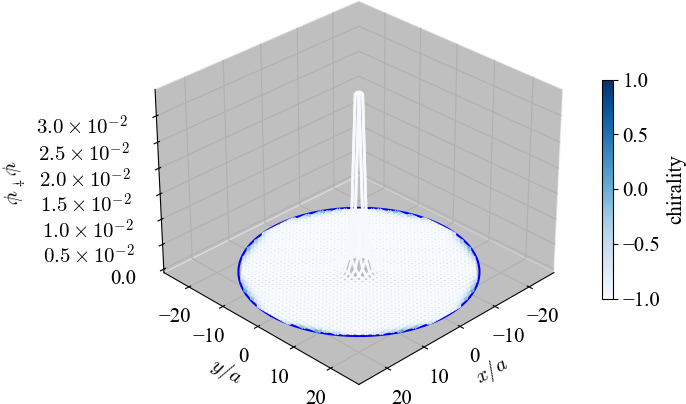}
  \caption{The amplitude of the zero mode at $z=a/2$ slice with $n=-1$.}
  \label{fig: eigenstate n=-1}
\end{figure}

Fig.~\ref{fig: spectrum n=1} shows the plot of the eigenvalue spectrum of $D_W^\dagger D_W$ for $n=1$ (upper panel) and $n=-1$ (lower panel).
The lattice setup is the same as in the previous section.
As in Fig.~\ref{fig: spectrum n=1}, the orange cross symbols represent the continuum results, while the filled circular symbols represent the numerical results on the lattice.
We see for $n=1$ that all modes below $mr_0$, including the zero mode, have chirality $\sigma_r=+1$, which is consistent with the continuum results.
IN contrast, for $n=-1$, we find a zero mode with chirality $\sigma_r=-1$, which does not appear in the continuum analysis.
The extra zero mode is localized at the center where the monopole is placed, as Fig~\ref{fig: eigenstate n=-1} shows.

\begin{figure}
  \centering
  \includegraphics[width=0.9\textwidth,bb= 0 0 354 291]{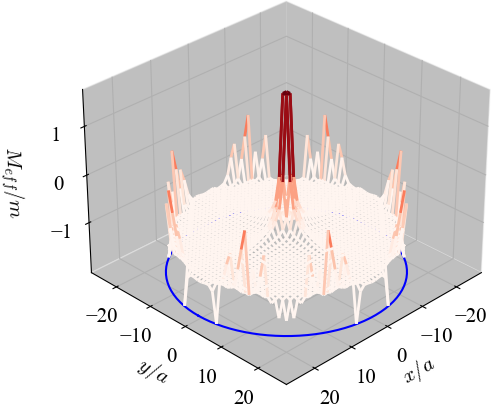}
    \caption{The plot of the effective mass defined in Eq.~\eqref{eq:eff_mass} for the center-localized zero mode at $z=a/2$ slice.}
    \label{fig: effective mass}
  \end{figure}

The origin of the zero mode localized at the center can be explained by the contribution to the effective mass from the Wilson term.
We define the effective mass as
\begin{align}
	M_{\rm eff}(x)=\frac{\phi_0^\dagger(x)\left(\sum^{3}_{i=1}{1\over 2a}\nabla_i\nabla_i^\dagger -m\right)\phi_0(x)}{\phi_0^\dagger(x)\phi_0(x)},
	\label{eq:eff_mass}
\end{align}
where $\phi_0(x)$ is the center-localized extra zero mode, and the term on the right-hand side proportional to $\nabla_i\nabla_i^\dagger$ represents the contribution from the Wilson term.
The numerical plot in the $z=1/2$ slice is provided in Fig.~\ref{fig: effective mass}.
We observe a peak at the center, where the sign of the effective mass flips.
We can interpret it as a domain-wall being generated in the vicinity of the monopole to populate the zero mode localized on it.
As a result, we cannot regard the low-energy effective theory of the system as a chiral gauge theory. 
Instead, the low-energy theory would actually be described by a vector-like gauge theory on the two domain walls.

\section{Summary}
In this article, we have studied the Shamir-type domain-wall fermion system with the spherical $S^2$ domain-wall.
The domain-wall is embedded in three-dimensional flat spacetime, where a single massive Dirac fermion resides.
When the gauge potential is trivial, as discussed in Section~\ref{sec: with gauge}, we have found that the almost chiral modes are localized on the $S^2$ domain-wall.
We have seen that the chiral modes feel gravity by analyzing the induced spin connection in the Dirac operator on the sphere.
We have also evaluated the finite volume effect and the restoration of the rotational symmetry in the continuum limit.
By turning on the nontrivial $U(1)$ gauge potential, the results change drastically.
We have found zero modes localized at the center, which possesses chirality opposite to that of the zero modes localized at the edge.
The centrally localized zero modes can be understood as arising from the domain-wall generated near the monopole.
The appearance of additional zero modes implies that the chiral gauge theory is not realized as a low-energy theory of this system.

\section*{Acknowledgements}
SA is supported by JSPS KAKENHI Grant Number JP23KJ1459.
HF is supported by JSPS KAKENHI Grant Number JP23K22490.
The work of NK is supported by JSPS KAKENHI Grant Number JP24KJ0157.

\bibliographystyle{JHEP}
\bibliography{ref.bib}

\providecommand{\href}[2]{#2}\begingroup\raggedright\begin{thebibliography}{10}

\bibitem{Kaplan:1992bt}
D.B.~Kaplan, \emph{{A Method for simulating chiral fermions on the lattice}}, \href{https://doi.org/10.1016/0370-2693(92)91112-M}{\emph{Phys. Lett. B} {\bfseries 288} (1992) 342} [\href{https://arxiv.org/abs/hep-lat/9206013}{{\ttfamily hep-lat/9206013}}].

\bibitem{Shamir:1993zy}
Y.~Shamir, \emph{{Chiral fermions from lattice boundaries}}, \href{https://doi.org/10.1016/0550-3213(93)90162-I}{\emph{Nucl. Phys. B} {\bfseries 406} (1993) 90} [\href{https://arxiv.org/abs/hep-lat/9303005}{{\ttfamily hep-lat/9303005}}].

\bibitem{Furman:1994ky}
V.~Furman and Y.~Shamir, \emph{{Axial symmetries in lattice QCD with Kaplan fermions}}, \href{https://doi.org/10.1016/0550-3213(95)00031-M}{\emph{Nucl. Phys. B} {\bfseries 439} (1995) 54} [\href{https://arxiv.org/abs/hep-lat/9405004}{{\ttfamily hep-lat/9405004}}].

\bibitem{Okumura:2016dsr}
K.-i.~Okumura and H.~Suzuki, \emph{{Fermion number anomaly with the fluffy mirror fermion}}, \href{https://doi.org/10.1093/ptep/ptw167}{\emph{PTEP} {\bfseries 2016} (2016) 123B07} [\href{https://arxiv.org/abs/1608.02217}{{\ttfamily 1608.02217}}].

\bibitem{Wen:2013ppa}
X.-G.~Wen, \emph{{A lattice non-perturbative definition of an SO(10) chiral gauge theory and its induced standard model}}, \href{https://doi.org/10.1088/0256-307X/30/11/111101}{\emph{Chin. Phys. Lett.} {\bfseries 30} (2013) 111101} [\href{https://arxiv.org/abs/1305.1045}{{\ttfamily 1305.1045}}].

\bibitem{Nash1956}
J.~Nash, \emph{The imbedding problem for riemannian manifolds}, {\emph{The Annals of Mathematics} {\bfseries 63} (1956) 20}.

\bibitem{Aoki:2022cwg}
S.~Aoki and H.~Fukaya, \emph{{Curved domain-wall fermions}}, \href{https://doi.org/10.1093/ptep/ptac075}{\emph{PTEP} {\bfseries 2022} (2022) 063B04} [\href{https://arxiv.org/abs/2203.03782}{{\ttfamily 2203.03782}}].

\bibitem{Aoki:2022aez}
S.~Aoki and H.~Fukaya, \emph{{Curved domain-wall fermion and its anomaly inflow}}, \href{https://doi.org/10.1093/ptep/ptad023}{\emph{PTEP} {\bfseries 2023} (2023) 033B05} [\href{https://arxiv.org/abs/2212.11583}{{\ttfamily 2212.11583}}].

\bibitem{Kaplan:2023pxd}
D.B.~Kaplan, \emph{{Chiral Gauge Theory at the Boundary between Topological Phases}}, \href{https://doi.org/10.1103/PhysRevLett.132.141603}{\emph{Phys. Rev. Lett.} {\bfseries 132} (2024) 141603} [\href{https://arxiv.org/abs/2312.01494}{{\ttfamily 2312.01494}}].

\bibitem{Kaplan:2023pvd}
D.B.~Kaplan and S.~Sen, \emph{{Weyl Fermions on a Finite Lattice}}, \href{https://doi.org/10.1103/PhysRevLett.132.141604}{\emph{Phys. Rev. Lett.} {\bfseries 132} (2024) 141604} [\href{https://arxiv.org/abs/2312.04012}{{\ttfamily 2312.04012}}].

\bibitem{Clancy:2024bjb}
M.~Clancy and D.B.~Kaplan, \emph{{Chiral edge states on spheres for lattice domain wall fermions}},  \href{https://arxiv.org/abs/2410.23065}{{\ttfamily 2410.23065}}.

\bibitem{Kaplan:2024ezz}
D.B.~Kaplan and S.~Sen, \emph{{Regulating chiral gauge theory at $\theta=0$}},  \href{https://arxiv.org/abs/2412.02024}{{\ttfamily 2412.02024}}.

\bibitem{Golterman:2024ccm}
M.~Golterman and Y.~Shamir, \emph{{Conserved currents in five-dimensional proposals for lattice chiral gauge theories}}, \href{https://doi.org/10.1103/PhysRevD.109.114519}{\emph{Phys. Rev. D} {\bfseries 109} (2024) 114519} [\href{https://arxiv.org/abs/2404.16372}{{\ttfamily 2404.16372}}].

\bibitem{Aoki:2024bwx}
S.~Aoki, H.~Fukaya and N.~Kan, \emph{{A Lattice Formulation of Weyl Fermions on a Single Curved Surface}}, \href{https://doi.org/10.1093/ptep/ptae041}{\emph{PTEP} {\bfseries 2024} (2024) 043B05} [\href{https://arxiv.org/abs/2402.09774}{{\ttfamily 2402.09774}}].

\bibitem{Aoki:2023lqp}
S.~Aoki, H.~Fukaya, N.~Kan, M.~Koshino and Y.~Matsuki, \emph{{Magnetic monopole becomes dyon in topological insulators}}, \href{https://doi.org/10.1103/PhysRevB.108.155104}{\emph{Phys. Rev. B} {\bfseries 108} (2023) 155104} [\href{https://arxiv.org/abs/2304.13954}{{\ttfamily 2304.13954}}].

\bibitem{Friedrich1980}
T.~Friedrich, \emph{Der erste eigenwert des dirac‐operators einer kompakten, riemannschen mannigfaltigkeit nichtnegativer skalarkr^^c3^^bcmmung}, \href{https://doi.org/10.1002/mana.19800970111}{\emph{Mathematische Nachrichten} {\bfseries 97} (1980) 117 }.

\bibitem{Atiyah:1963zz}
M.F.~Atiyah and I.M.~Singer, \emph{{The index of elliptic operators on compact manifolds}}, \href{https://doi.org/10.1090/S0002-9904-1963-10957-X}{\emph{Bull. Am. Math. Soc.} {\bfseries 69} (1969) 422}.

\end{thebibliography}\endgroup

\end{document}